**Comment on "Identification of Different Electron Screening Behavior Between the Bulk and Surface of (Ga,Mn)As"**

In a recent Letter [1], Fujii *et al.* reported Mn 2*p* photoelectron emission spectra for (Ga,Mn)As recorded using hard x-rays. Due to the enhanced bulk sensitivity, hard-x-ray spectra reveal an extra "low-binding-energy peak", which is absent in surface-sensitive spectra recorded using soft x-rays. Based on Anderson-impurity-model calculations, Fujii *et al.* assigned the low-binding-energy peak to a $\underline{c}d^6\underline{L}^2$ ($\underline{c}d^6\underline{h}^2$) final state, and related the variations in its intensity to variations in the As 4*p*-Mn 3*d* hybridization strength *V*. We show here that the definition of the charge-transfer energy $\Delta$ considered by Fujii *et al.* is different from that considered in the Zaanen-Sawatzky-Allen (ZSA) diagram [2]. We note that the Anderson impurity model is insufficient to describe low-binding-energy peaks in hard-x-ray core-level photoemission for transition-metal (TM) compounds on the verge of a metal-insulator transition. We propose a more plausible origin for the (Ga,Mn)As low-binding-energy peak, related to the nature of its metal-insulator transition.

(i) *Location of (Ga,Mn)As in the ZSA diagram.*—The charge-transfer energy $\Delta$ is defined [2] as the energy required to transfer an electron from the ligand *p* to the TM *d* orbitals $\Delta \equiv E(d^{n+1}\underline{L}) - E(d^n)$. Here, $d^n$ is the "perfectly ionic configuration" for the neutral *n*-electron system (initial state), corresponding to ligands with *closed-shell* configuration ($p^6$). This definition of $\Delta$ applies also to "negative-$\Delta$ compounds", which have a $d^{n+1}\underline{L}$-like charge-transfer-type ground state [3,4]. The ionic $d^n$ level is used as energy reference to construct the ZSA diagram. The perfectly ionic configuration for Mn in (Ga,Mn)As is $d^4$ (Mn$^{3+}$ formal valence). Hence, to locate (Ga,Mn)As in the ZSA diagram, the charge-transfer energy to be considered is $\Delta(\text{ZSA}) = E(d^5\underline{L}) - E(d^4)$, which differs from the definition used by Fujii et al $\Delta(\text{Fujii}) = E(d^6\underline{L}^2) - E(d^5\underline{L})$ by the intra-atomic Coulomb-interaction energy $\Delta(\text{ZSA}) = \Delta(\text{Fujii}) - U$. Note that $E(d^n \underline{L}^m) = n\varepsilon_d^0 - m\varepsilon_p + \frac{1}{2}n(n-1)U$ (Ref. [5]). Taking $U = 4\,\text{eV}$, the $\Delta$-value assigned to (Ga,Mn)As by Fujii *et al.* $\Delta(\text{Fujii}) = 1\,\text{eV}$ corresponds to a *negative* value in the ZSA-diagram scale $\Delta(\text{ZSA}) = -3\,\text{eV}$. This result is similar to that of Okabayashi *et al.* in their ($Mn^{3+}$) analysis of soft-x-ray spectra [6]. The natures of the ground state, conductivity gap, and metal-insulator transition for (Ga,Mn)As are thus understood if classified as a negative-$\Delta$ compound.

(ii) *Sensitivity of "low-binding-energy peaks" to d-d interactions.*—Low-binding-energy peaks, observed in core-level photoemission for TM compounds near a metal-insulator transition, were shown to correspond to final states reached through interactions between two or more TM sites (nonlocal screening) [7,8]. They indicate the onset of metallicity. Since the Anderson impurity model only accounts for a single TM site, it cannot describe low-binding-energy peaks resulting from intersite *d-d* interactions. Multisite cluster models [7,8], which account for the possibility of intersite interactions, point to a different origin for the extra peak in hard-x-ray (Ga,Mn)As photoemission. The extra (Ga,Mn)As peak plausibly arises from nonlocal screening, involving charge-excitations across a nearly zero conductivity gap. As shown by Taguchi *et al.* for other TM compounds [9,10], intensity variations of the (Ga,Mn)As low-binding-energy peak rather relate to variations in the *d-d* hybridization strength $V^*$.

(iii) *Nature of the (Ga,Mn)As metal-insulator transition. Peak assignment.*—As a negative-$\Delta$ compound with a $d^5\underline{L}$-like ground state, the smallest-energy charge excitations for (Ga,Mn)As should be of the type $(d^5\underline{L})_i + (d^5\underline{L})_j \rightarrow (d^5)_i + (d^5\underline{L}^2)_j$ (Ref. [3,4]). The (Ga,Mn)As insulator-to-metal transition thus corresponds to the closing of a gap of *p-p* type. Conduction occurs by hopping of *p*-holes from the vicinity of one Mn site to that of another. In this context, the extra (Ga,Mn)As peak at ~638.5 eV, rather than originating from local (*p*-band) screening ($\underline{c}d^6\underline{L}^2$), corresponds to a $\underline{c}d^5$ final state, reached by "ejecting" the ground-state ligand hole ($\underline{L}$) to the vicinity of another Mn site ($d^5\underline{L}^2$). This mechanism is promoted by Coulomb repulsion between the Mn 2$p$ core hole ($\underline{c}$) and the As 4$p$ ligand hole ($\underline{L}$) at the photoemission site. Consistent with the previous ($Mn^{3+}$) analysis of Okabayashi *et al.* [6] and with the level ordering claimed by Fujii *et al.*, the peak at ~640 eV should essentially correspond to a locally ($p$-band) screened $\underline{c}d^6\underline{L}^2$ final state, while the broad high-binding-energy satellite at ~643 eV should essentially correspond to a $\underline{c}d^5\underline{L}$ final state. The absence of low-binding-energy peak in surface-sensitive soft-x-ray (Ga,Mn)As photoemission is understood by taking into account that downward band bending causes hole depletion close to the surface. In the absence of conducting holes, the nonlocal screening mechanism is not active.

M. Moreno

    Instituto de Ciencia de Materiales de Madrid (CSIC)

    Cantoblanco, E-28049 Madrid, Spain

K. H. Ploog

    Paul-Drude-Institut für Festkörperelektronik

    D-10117 Berlin, Germany